\documentclass[11pt]{article}
\usepackage{moriond,epsfig}
\usepackage{subfig}
\usepackage{tikz}
\usepackage{url}

\newcommand{\met}{\mbox{$\protect \raisebox{0.26ex}{$\not$}E_T$}}
\newcommand{\mpt}{\mbox{$\protect \raisebox{0.26ex}{$\not$}p_T$}}
\newcommand{\mlep}{\mbox{$\protect \raisebox{0.26ex}{$\not$}\ell$}}
\newcommand{\invfb}{\mbox{fb$^{-1}$}}
\newcommand{\gevcc}{\ensuremath{\mbox{~GeV}/c^2}}
\begin{document}
\vspace*{4cm}
\title{Search for rare SM processes in the \met+$b$-jets signature at CDF}

\author{ K. Potamianos }

\address{Purdue University, West Lafayette, IN, USA}

\maketitle\abstracts{
The missing transverse energy (\met) plus b-jets signature is very promising for searches for the Higgs boson or new physics. Indeed, \met\ naturally arises from unidentified particles such as neutrinos, neutralinos, gravitons, etc., and b-quarks are the main decay products of a low mass Higgs boson as well as of several exotic particles.
The main challenge is to identify and reject the numerous standard model (SM) backgrounds that mimic this signature. This is especially so for QCD multi-jet production, a large background due to mis-measurement (rather than undetectable particles).
We present state-of-the-art data-driven and multivariate techniques to characterize and reject this instrumental background. These techniques make analyses in this signature as sensitive as those using lepton identification and allow probing for rare SM processes.
We describe searches for electroweak single top production, a part of the observation of single top by CDF, and for a low mass SM Higgs boson, one of the most sensitive among low mass Higgs searches at CDF. We also present a measurement of the top pair cross-section in this signature, and discuss other analyses and future prospects.
}

\section{Introduction}

The standard model of particle physics (SM) accurately describes most physical observables not involving gravity. All of its particles have been observed and characterized, save to the Higgs boson~\cite{smHiggs}, whose role is to provide mass to the elementary particles.
Direct searches and precise electroweak fits constrain its mass, and favor a low mass ($m_H<158\gevcc$ at 95 \% C.L.)~\cite{LEPEWWG}. 

At the Tevatron, the low mass range is best investigated through the associated production of a Higgs boson and a $W$ or $Z$ boson, whose leptonic decay products are triggered on~\cite{Potamianos:2010vp}. At low mass, the Higgs predominantly decays to a $b\bar{b}$ pair. In the following, we focus on $ZH\to\nu\nu b\bar{b}$ events.
 These yield a large transverse energy imbalance (\met), two high-$p_T$ $b$-jets, and no identified lepton. Due to the difficulty to identify and reconstruct leptons, we are also sensitive to $WH\to\mlep\nu b\bar{b}$. This is true especially when the $W$ boson decays to a $\tau$ that is not reconstructed ($\sim50\%$ of the times). Finally, we accept a tiny fraction of $ZH\to\mlep\mlep b\bar{b}$ events.

In addition to Higgs physics, this signature is also sensitive to the electroweak production of a single top quark, diboson production, and top pair production. These SM processes are either categorized as signal or background, while events from QCD multi-jet production and from the production of a vector boson in association with jets are part of the background. 

Next to events from SM processes, many models of new physics also predict this signature: SUSY ($\tilde{b}\bar{\tilde{b}}\rightarrow b\tilde{\chi^0}\bar{b}\tilde{\chi^0}$), technicolor ($\rho^\pm_T \rightarrow Z\pi^\pm_T \rightarrow \nu\nu b\bar{q}$ and $\rho^\pm_T \rightarrow W^\pm \pi^0_T \rightarrow \ell\nu b\bar{b}$), extra-dimension, etc. In this proceeding, we focus only on SM physics.

In the \met+$b$-jets signature, the signal is very small when compared to the large backgrounds, especially from QCD production.
In the following, we present a innovative technique to isolate and reject the large QCD background. 

\section{Background modeling}

The \met+$b$-jets signature accepts events from every SM process.
We use a combination Monte Carlo and data-driven techniques to provide a proper model for each of them.
The electroweak processes yield \emph{real} \met\ coming from the neutrino(s) in the event. We use Monte Carlo to model these~\cite{Aaltonen:2010fs}. Additionally, we use data to predict the rate at which light flavor events from these processes can be wrongly identified as originating from a $b$-quark (i.e. $b$-tagged).
On the contrary, QCD multi-jet production (MJ) yields \emph{instrumental} \met, due to the mis-measurement of the energy of the jets. Because of a large cross-section and of the presence of non-negligible high order effects, it is impractical and not advised to use Monte Carlo to model MJ, especially with large datasets. We therefore use a data-driven model that defines a four-dimensional matrix to predict the probability for a data event to be $b$-tagged: the \emph{Tag-Rate-Matrix}~\cite{Aaltonen:2010fs}.

\vspace*{-2mm}
\section{Tools for background rejection}

We select events with $\met>50\gevcc$, two or three jets ($>15\gevcc$, $|\eta|<2$), and no identified lepton. We further require the leading (second) jet to have an energy of at least 35 (25)\gevcc\ and either one to be central ($|\eta|<0.9$). We remove mis-measured QCD events requiring $\Delta\phi(\met,j_{(2,3)})>0.4$ and $\Delta\phi(\met,j_1)>1.5$. We then further improve the S/B ratio.

\vspace*{-2mm}
\subsection{Signature of missing particles}

A common technique to identify a particle such a neutrino is to measure the transverse energy imbalance in the calorimeter (\met). Here, we also rely on an \emph{independent} sub-detector, the spectrometer, to determine the transverse momentum flow imbalance (\mpt). 
In the case of a particle escaping the detector, the \met\ and the \mpt\ are aligned. However, for a mis-measured QCD di-jet event, the \met\ and is either aligned or anti-aligned to the \mpt\ (Figure~\ref{fig:dPhiMET_MPT}, left). 

\providecommand{\plotWidth}{3cm}
\begin{figure}[t]
\providecommand{\plottype}{MethodAM_ShapeComp}
\begin{center}
\hspace*{-0.75cm}
\begin{tabular}[t]{cc}
\begin{minipage}{12cm}

\subfloat[]{\label{fig:NNinputs_shapeComp}\label{fig:dPhiMET_MPT}
\includegraphics[width=\plotWidth]{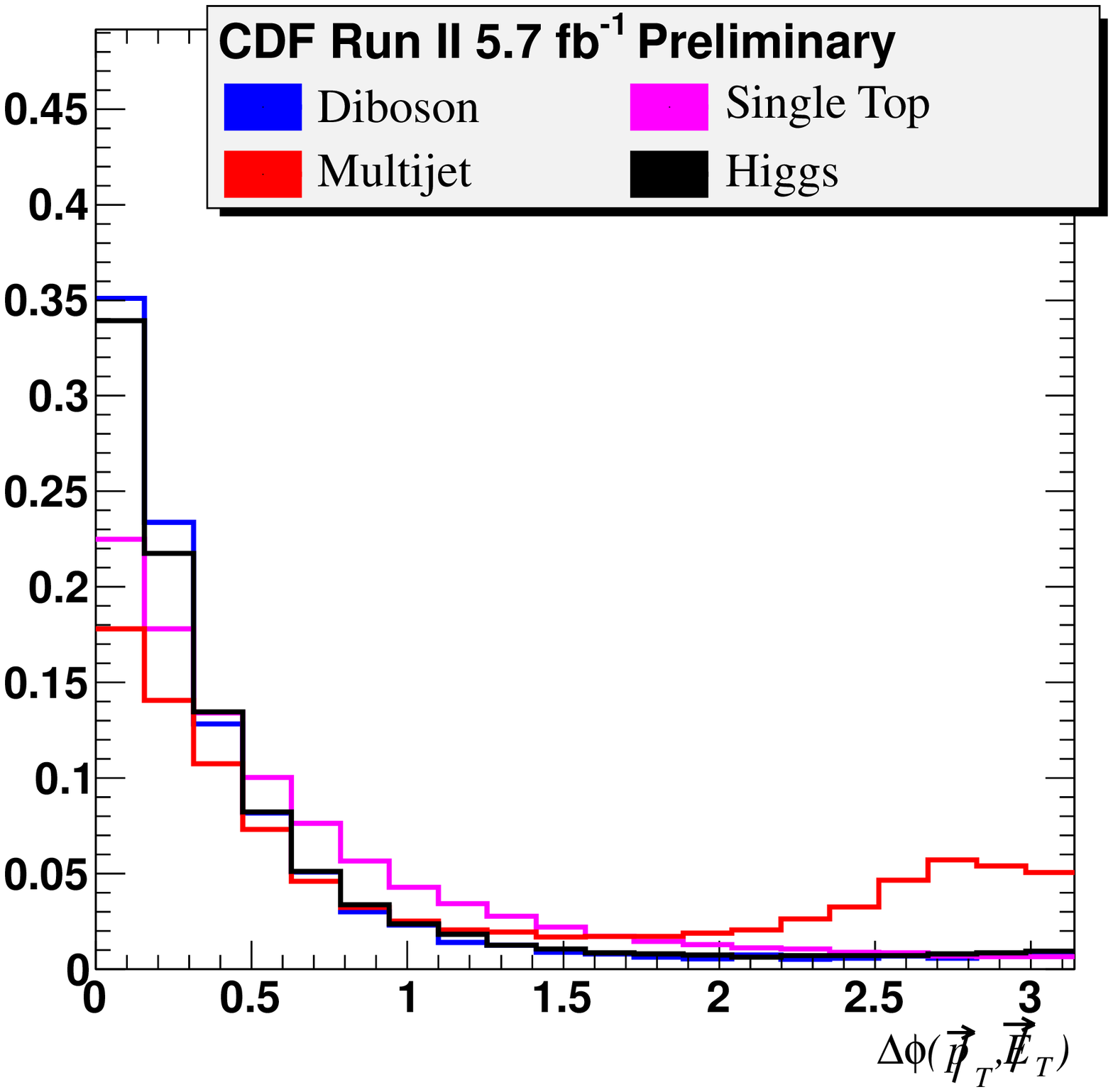}
\includegraphics[width=\plotWidth]{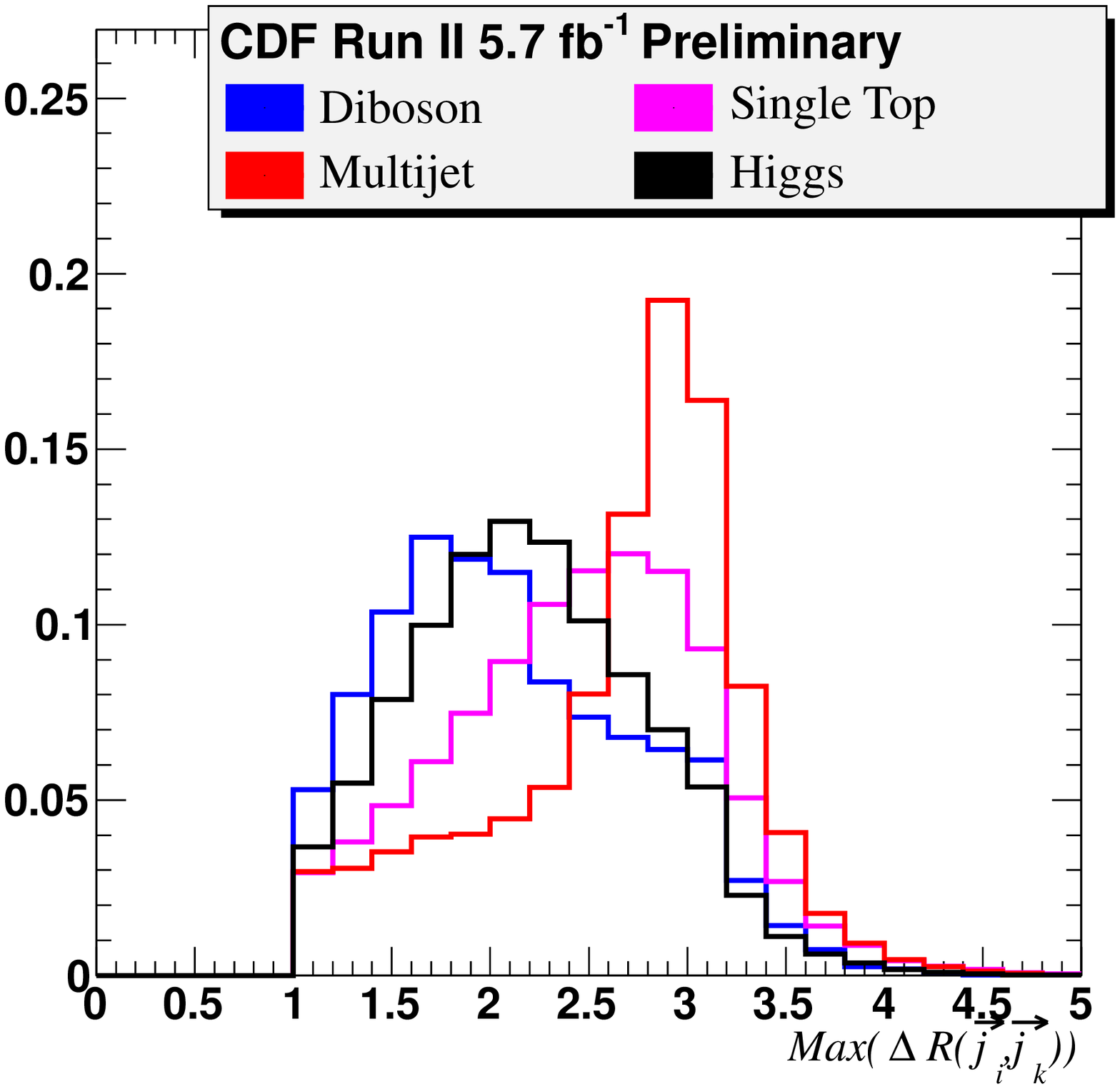}
\includegraphics[width=\plotWidth]{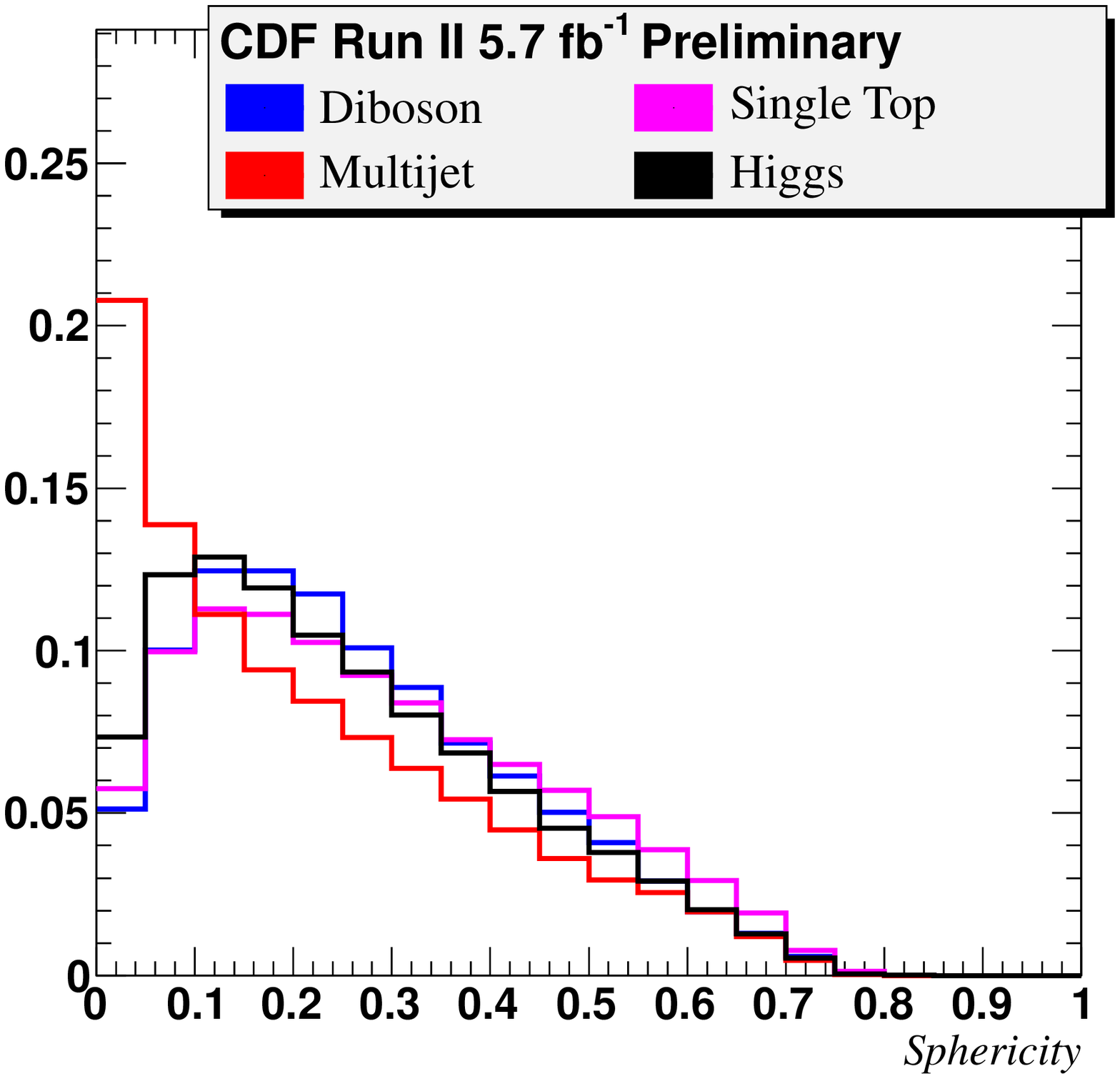}
\includegraphics[width=\plotWidth]{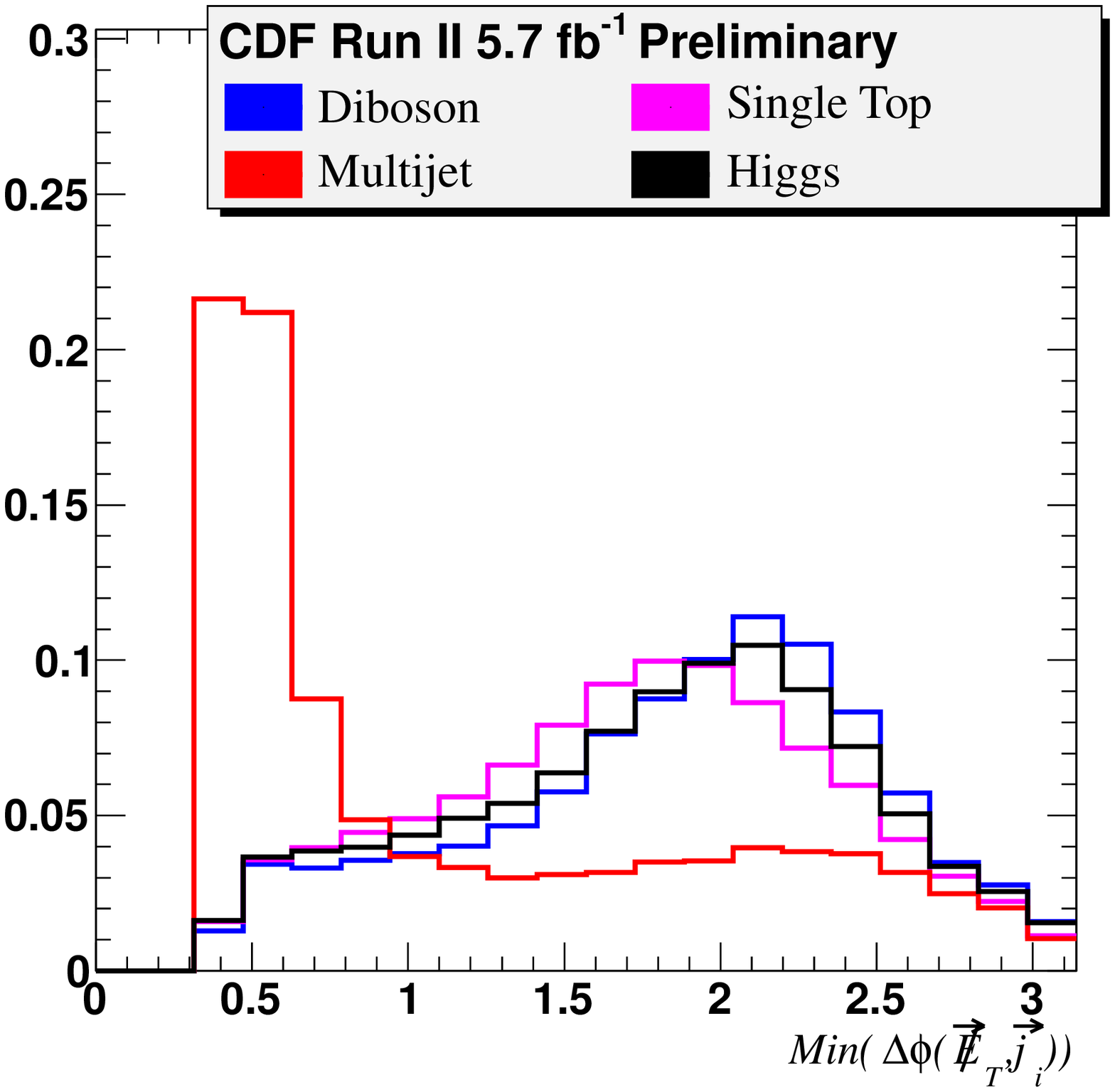}}

\renewcommand{\plottype}{MethodAM}
\subfloat[]{\label{fig:NNinputs_validationPlot}
\includegraphics[width=\plotWidth]{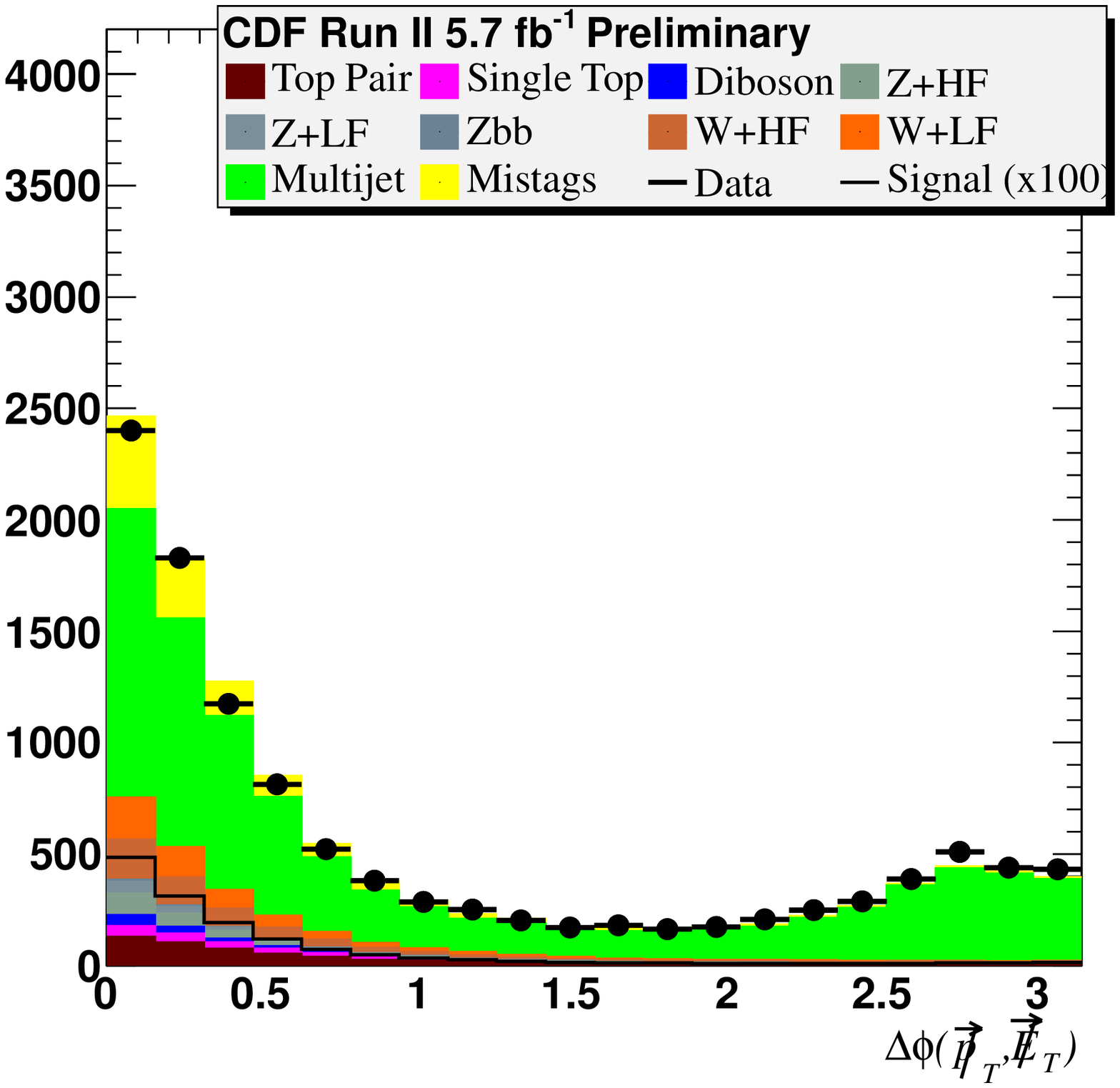}
\includegraphics[width=\plotWidth]{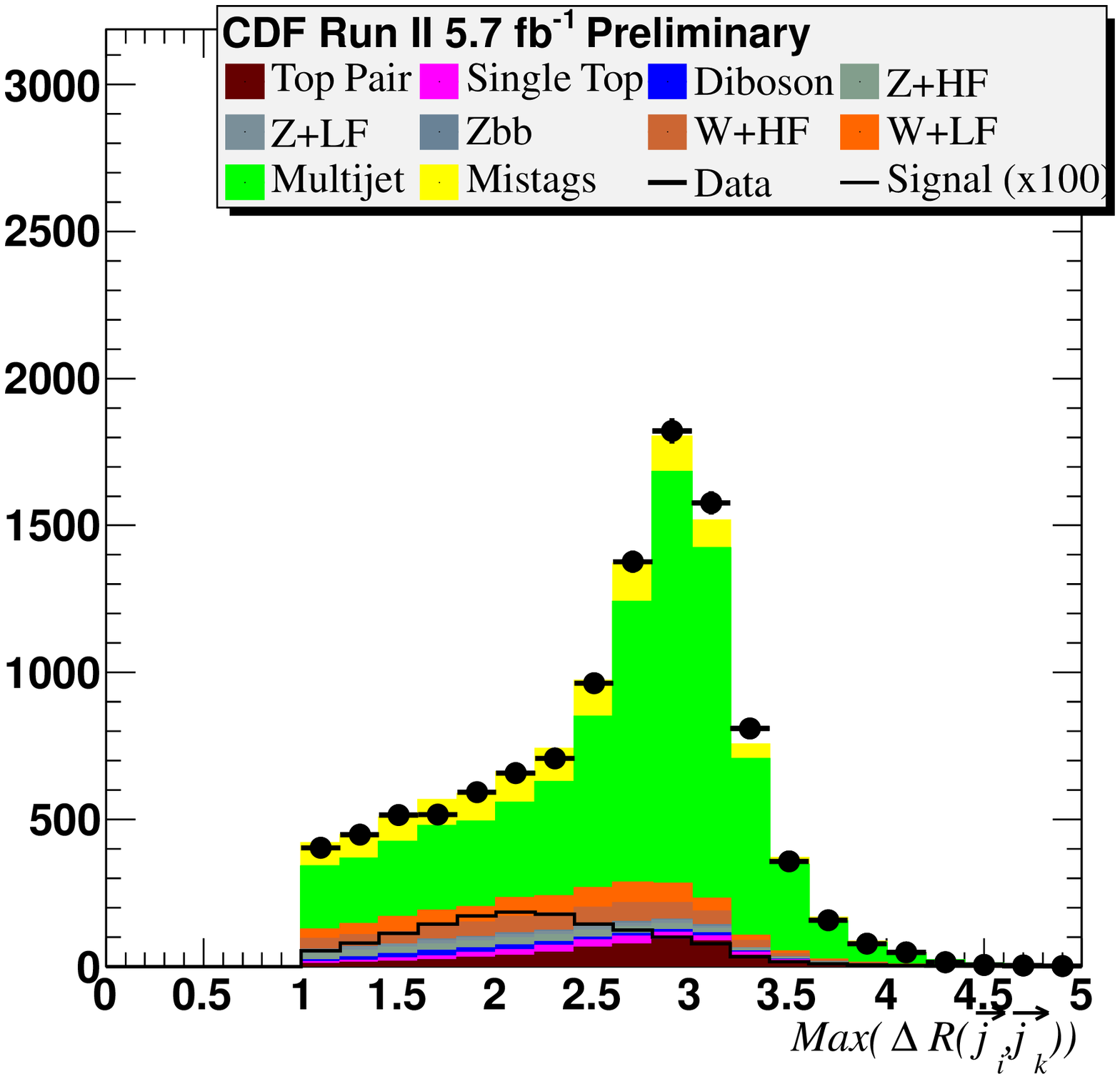}
\includegraphics[width=\plotWidth]{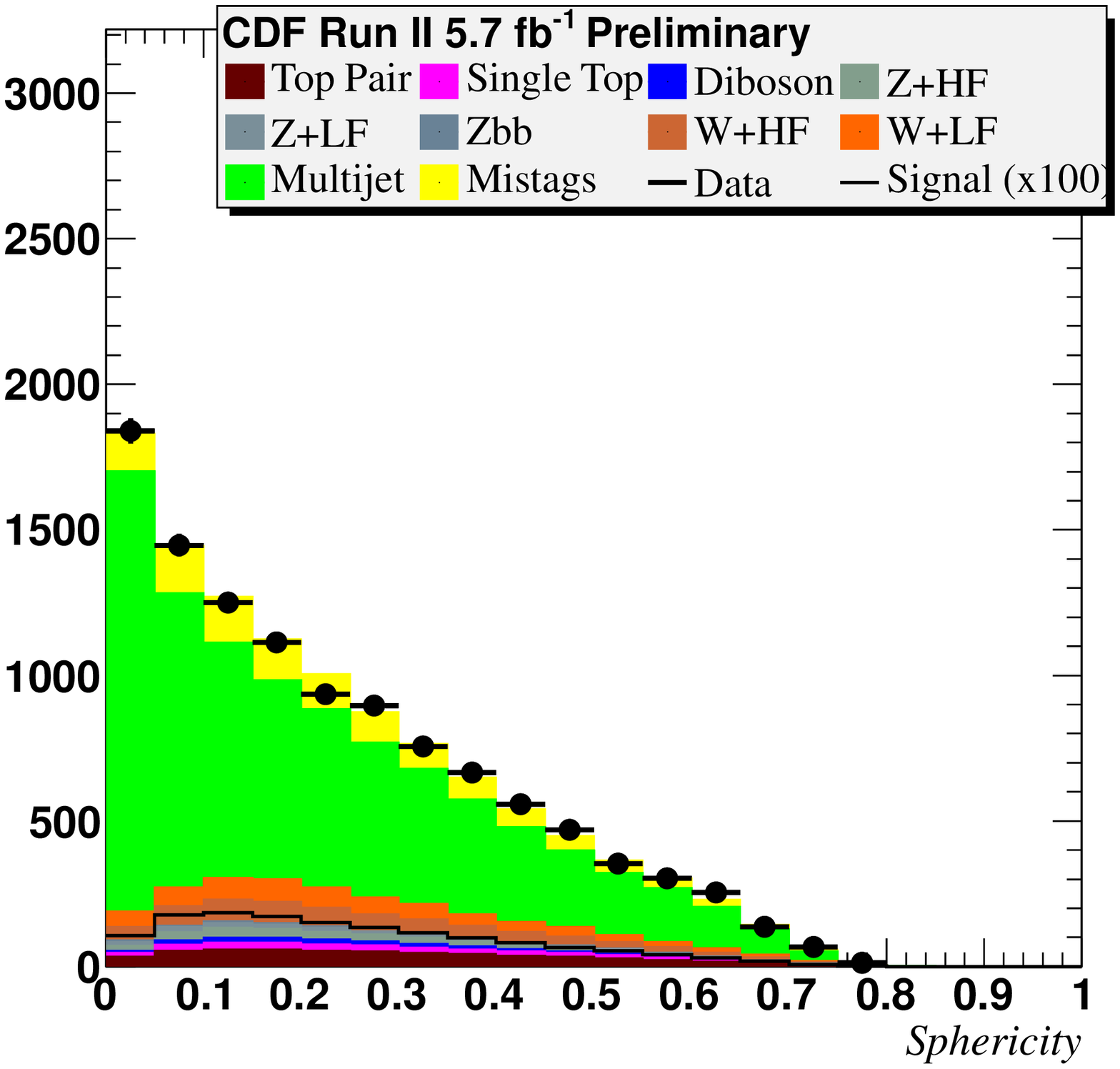}
\includegraphics[width=\plotWidth]{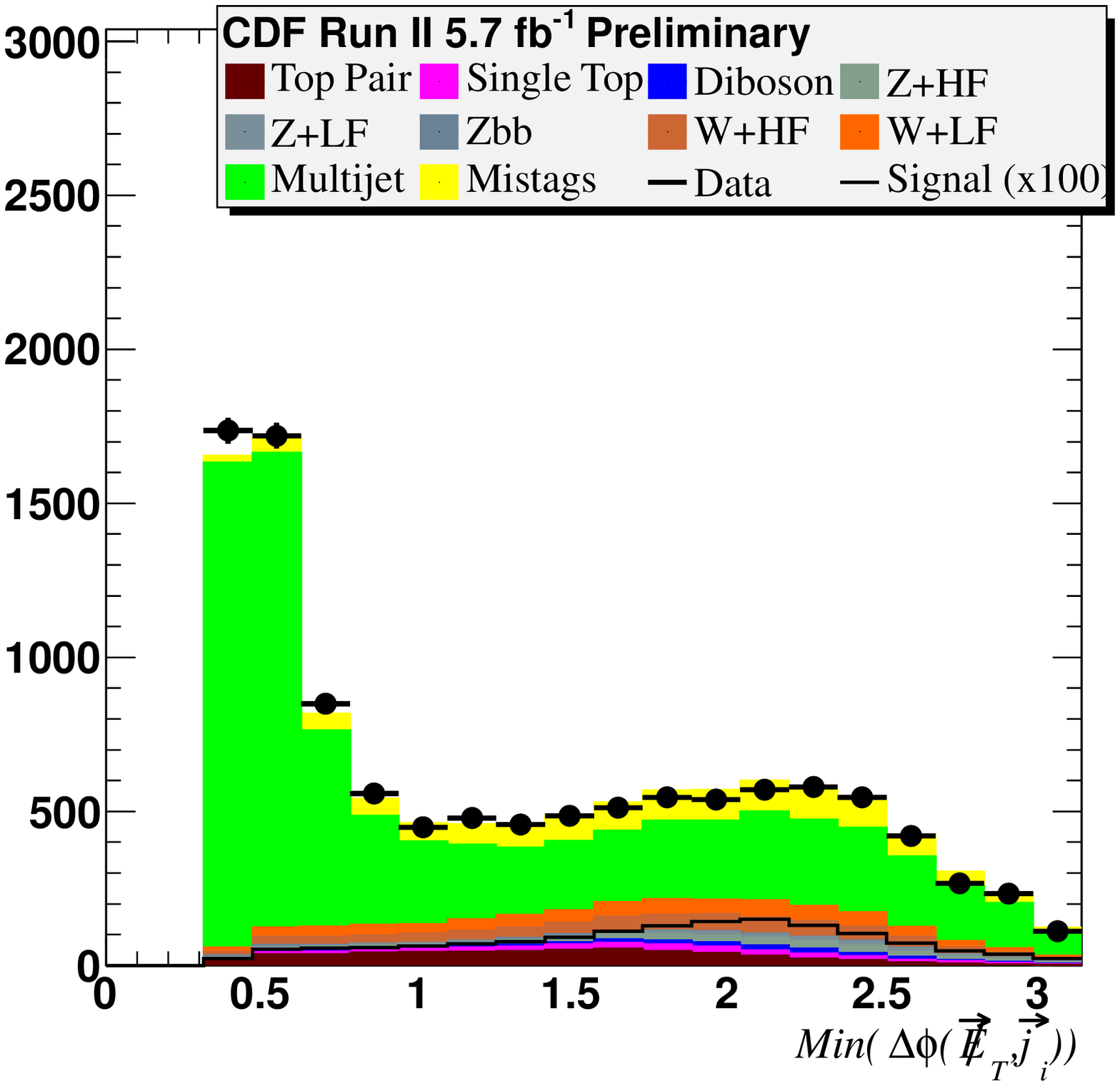}}

\end{minipage}
&

\begin{minipage}{3cm}

\subfloat[]{
\includegraphics[width=\plotWidth]{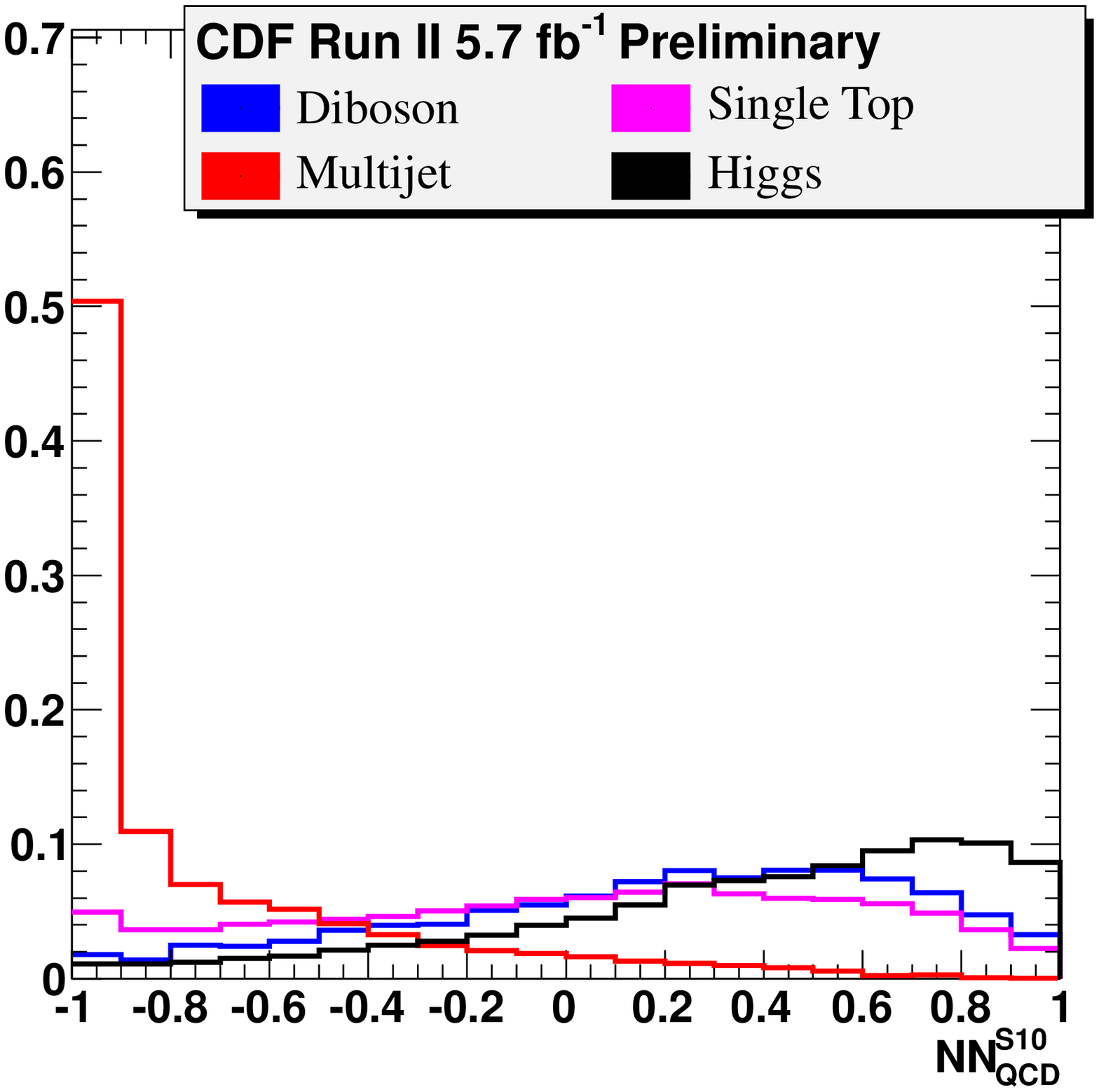}
}

\subfloat[]{
\includegraphics[width=\plotWidth]{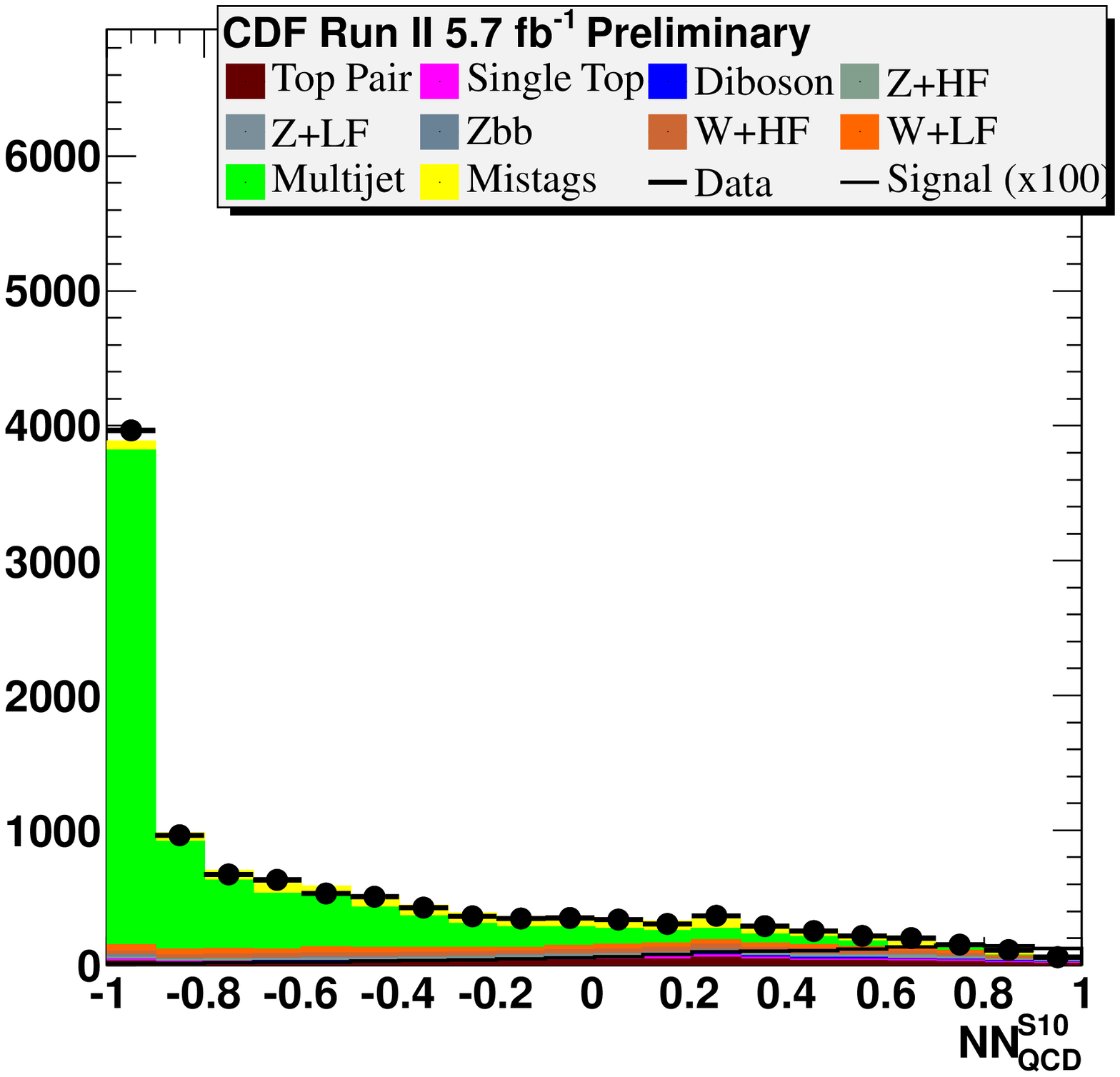}
}

\end{minipage}

\end{tabular}

\end{center}

\caption{Some of the input variables to the neural network used to reject the QCD multi-jet background: (a) shape comparison, and (b) model validation. (c), (d): Neural network output.}
\label{fig:NNinputs}
\end{figure}

\vspace*{-2mm}
\subsection{A neural network to reject the QCD multi-jet background}

Next to comparing \mpt\ to \met, we identify kinematic quantities discriminating the signal from the QCD background.
Instead of cutting on each variable, we feed a neural network that exploits their correlations, and then cut on its output, reducing the background more efficiently.

We train our network using the \emph{pre-tag} data sample weighted by the \emph{Tag-Rate-Matrix} to model MJ events. This choice is preferred over the use of a Monte Carlo QCD background because it allows not only to reject QCD but also a part of the other electroweak backgrounds. With this technique, it is possible to reject about 90\% (70\%) of the MJ (overall) background while only loosing 10\% of the signal. The signal sample is composed of Monte Carlo events in proportion to the relative size of each signal component.

Figure~\ref{fig:NNinputs_shapeComp} shows the difference in shape for some of the network input variables. 
Each of the variables is validated against data, as shown in Figure~\ref{fig:NNinputs_validationPlot}. In addition to the region where the network is derived, we use five other control regions in which we check both the inputs to the network and its output. This makes us confident in using multivariate analysis to find a signal. 

\section{Analyses using this neural network background rejection technique}

The technique we present is very generic and can be trained with different signals. We briefly describe several analyses that owe their significance to this technique. After rejecting the QCD background, we train another neural network to discriminate the signal from the background.

\vspace*{-2mm}
\subsection{Measurement of the top pair production cross-section}

First observed in 1995 \cite{Abe:1995hr}, the top quark has been extensively studied at the Tevatron, mostly in the (semi-)leptonic and all-hadronic signatures. In 2010, we measured for the first time the top pair production cross-section in the \met+$b$-jets signature~\cite{cdfNote:10237}. This measurement is complementary to those preceding and contributes to improving the world average. Much importantly, probing a well known signal is a stringent test of the analysis technique. Analyzing 5.7\invfb\ of CDF data, we measure a cross-section of $7.12^{+1.20}_{-1.12}$ pb, assuming $m_t=172.5\gevcc$. This measurement is as sensitive as the dilepton and all-hadronic signatures (Figure~\ref{fig:ttbarMETbb}).

\vspace*{-2mm}
\subsection{Measurement of the single top production cross-section}

In 2009, CDF and DZero observed the electroweak production of a single top quark~\cite{Aaltonen:2009jj}. The single top production cross-section is directly proportional to the square of the $|V_{tb}|$ element of the CKM matrix~\cite{CKM} and a measurement thus constrains fourth-generation models.  Our analysis in the \met+$b$-jets signature~\cite{Aaltonen:2010fs} contributed to the result by adding 30\% orthogonal signal into the combination~\cite{Aaltonen:2009jj,Aaltonen:2010jr}. Analyzing 2.1\invfb\ of CDF data, we measure a cross-section of $4.9^{+2.5}_{-2.2}$ pb (observed sensitivity of 2.1$\sigma$) and a $|V_{tb}|$ value of $1.24^{+0.34}_{-0.29} \pm 0.07$ (theory), assuming $m_t=175\gevcc$ (Figure~\ref{fig:singleTopMETbb}).

\vspace*{-2mm}
\subsection{Search for the SM Higgs boson}

\begin{figure}[t]
\begin{center}
\subfloat[Top pair production]{\label{fig:ttbarMETbb}
\begin{tikzpicture}
\draw[white,fill=white] (0,0) rectangle (5,5);
\node [inner sep=0pt,above right] at (0.06,0.04) {\includegraphics[width=4.9cm]{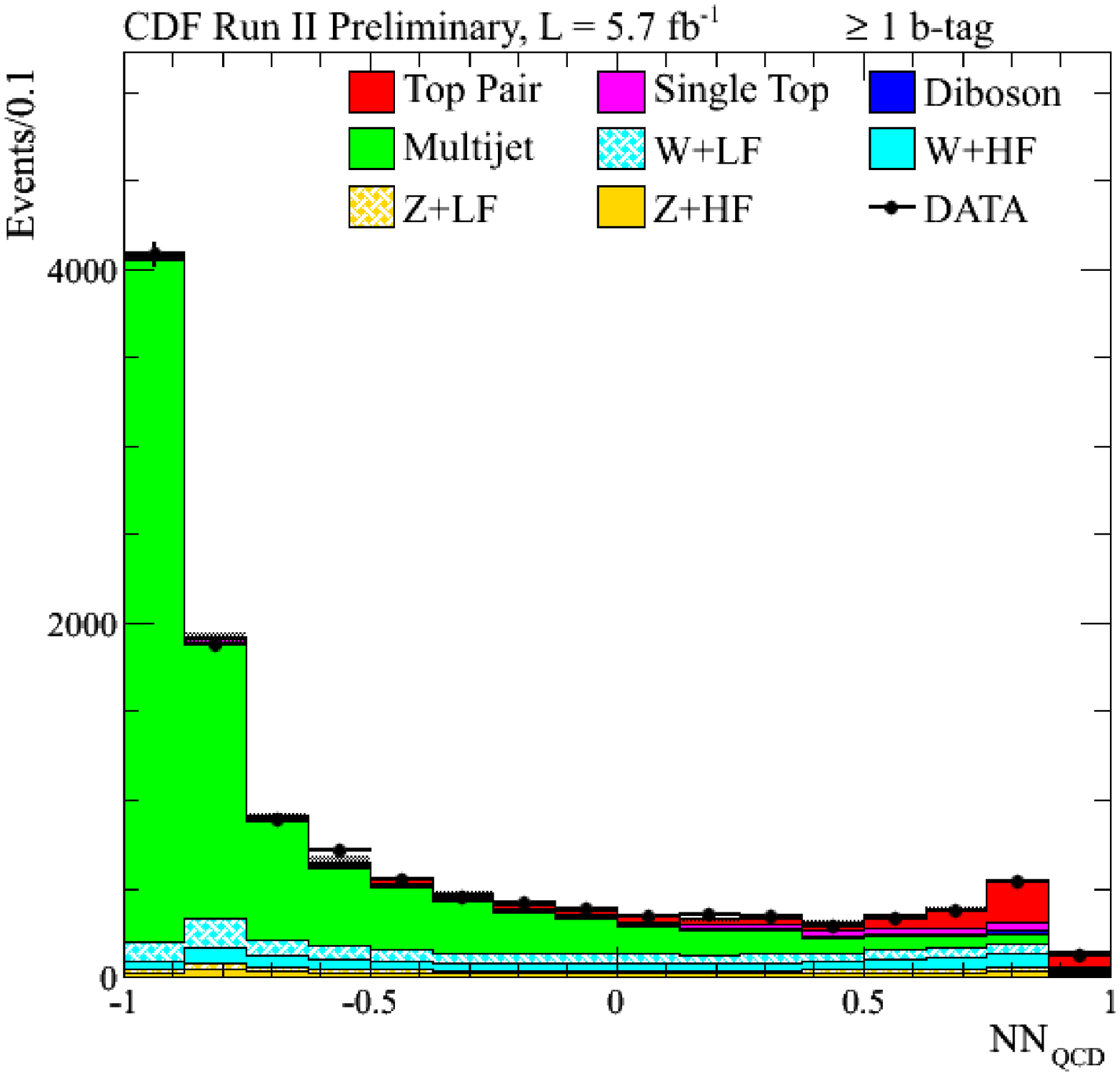}};
\node [inner sep=0pt,above right] at (1.5,1.3) {\includegraphics[width=3.3cm]{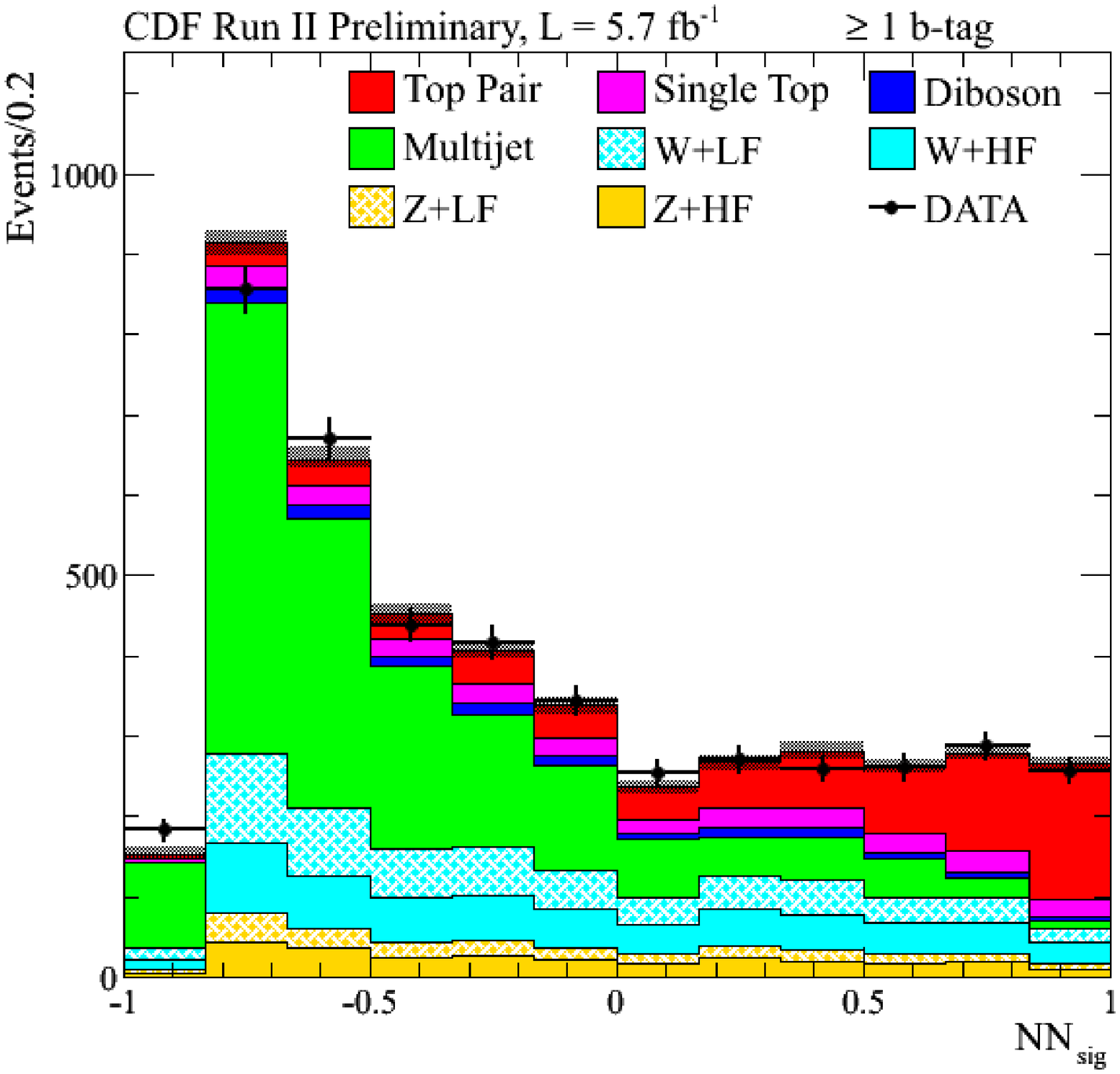}};
\draw [red, thick] (1.7,0.4) -- (1.7,1.6);
\draw [red,thick] (3.4,0.7) ellipse (1.7 and .5);
\draw [red, ultra thick,->] (3.5,0.9)  -- (3.5,1.5) ;

\node [inner sep=0pt,above right] at (0,-4) {\includegraphics[width=4.9cm]{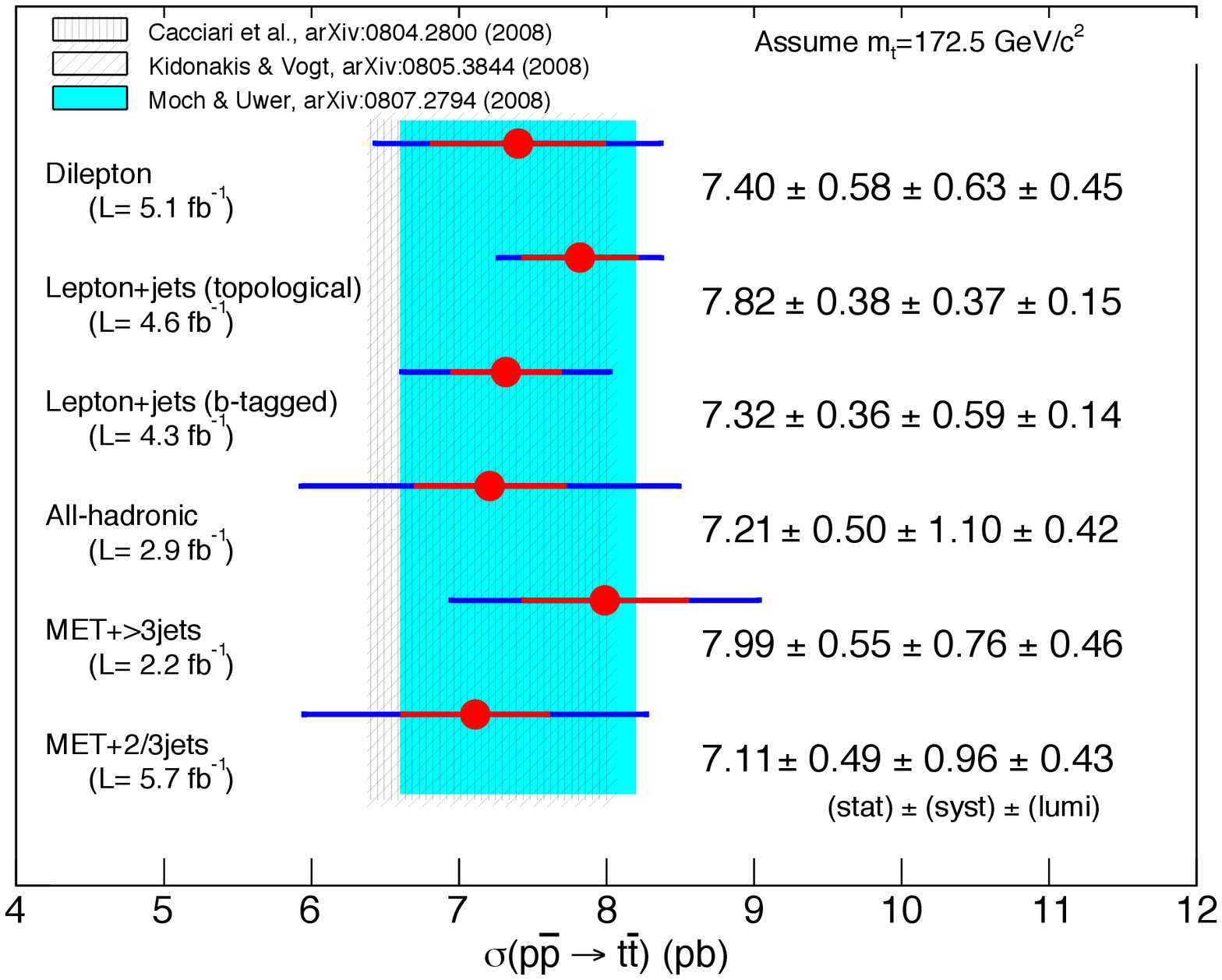}};
\draw [red,thick] (.1,-3.45) rectangle (4.6,-2.85);
\end{tikzpicture} 
}
\subfloat[EW single top]{\label{fig:singleTopMETbb}
\begin{tikzpicture}
\draw[white,fill=white] (0,0) rectangle (5,5);
\node [inner sep=0pt,above right] at (0,0) {\includegraphics[width=5cm]{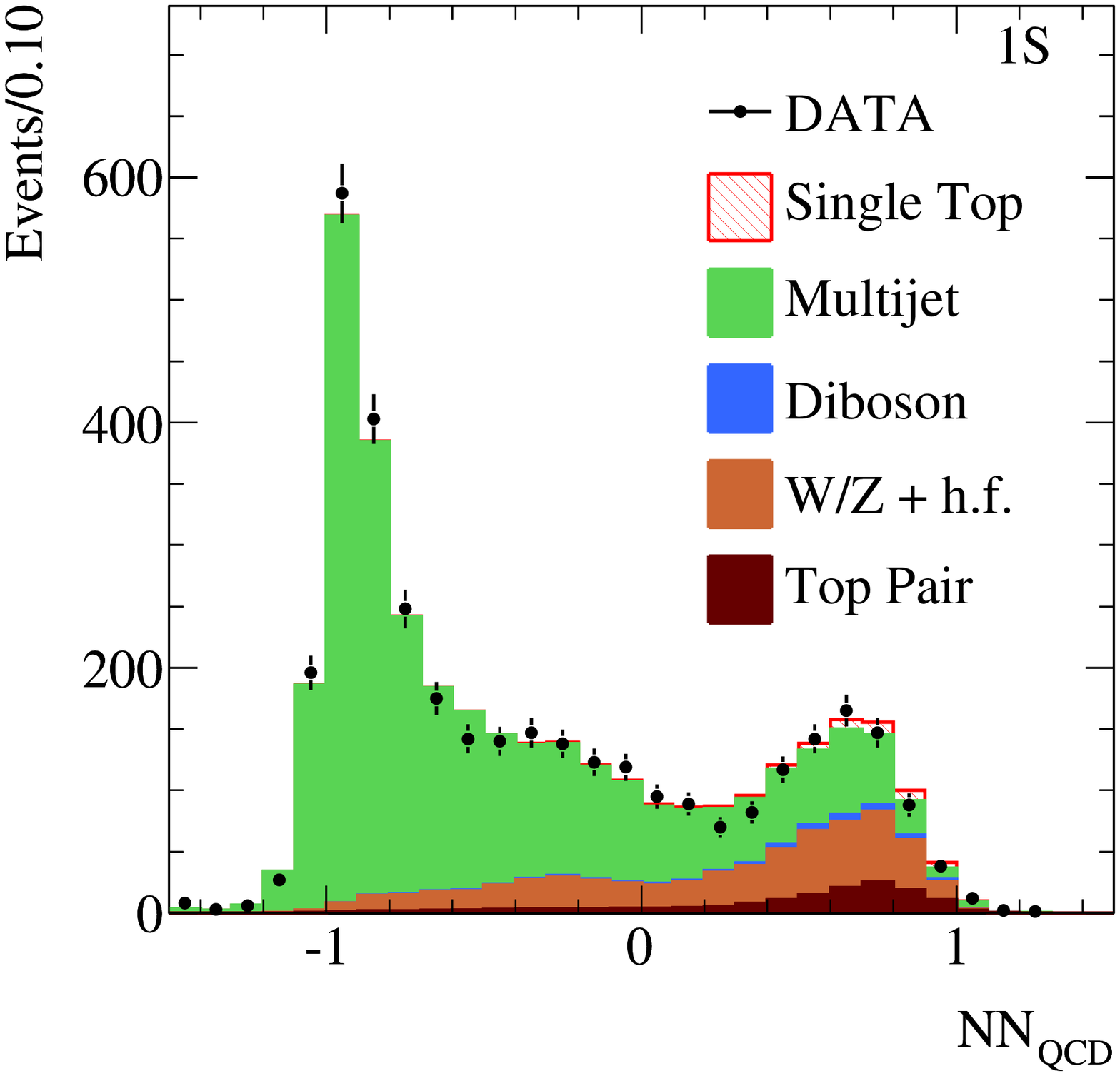}};
\node [inner sep=0pt,above right] at (1.9,1.8) {\includegraphics[width=3cm]{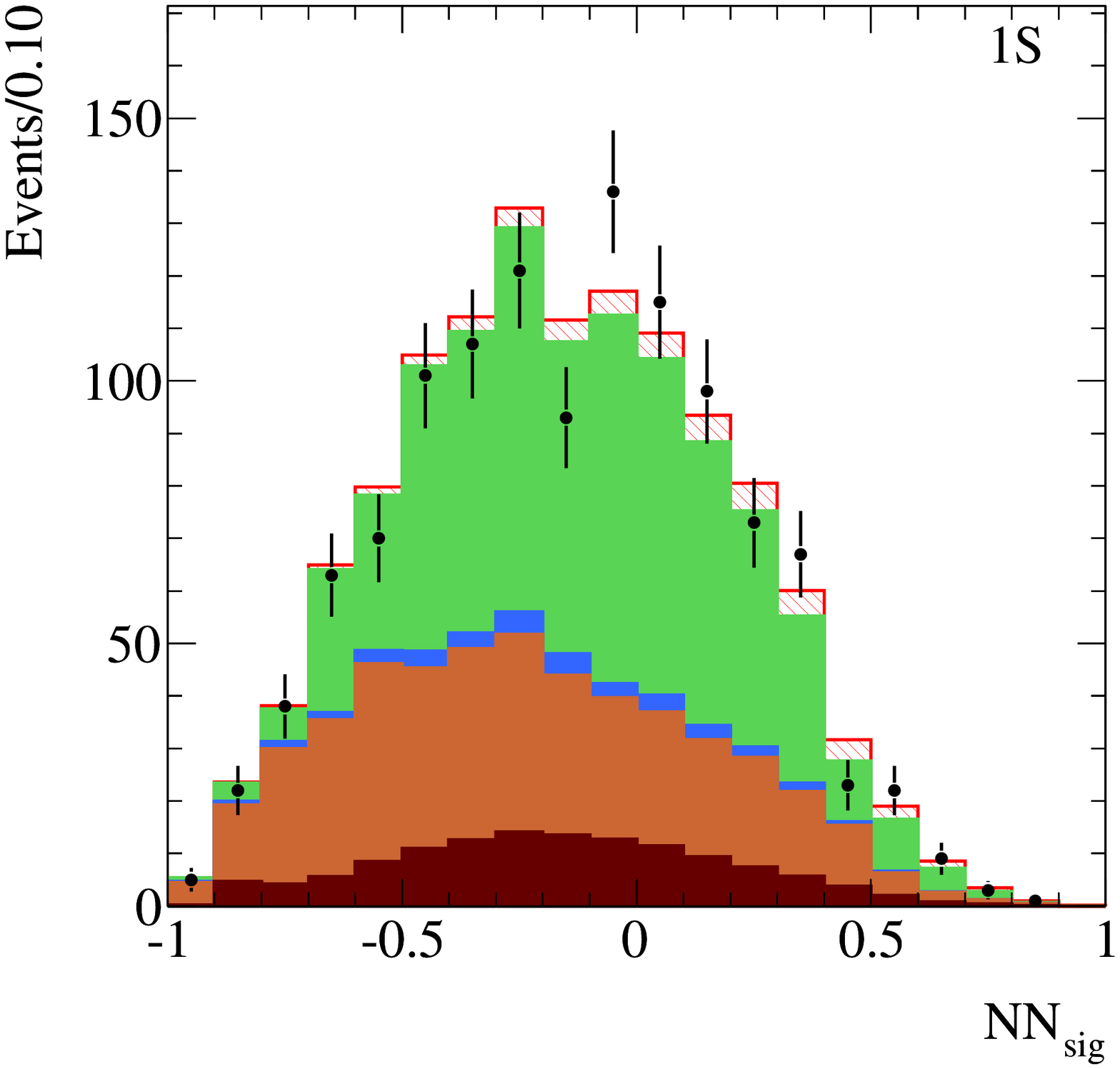}};
\draw [red, thick] (2.65,0.7) -- (2.65,1.9);
\draw [red,thick] (3.8,1.2) ellipse (1.15 and .5);
\draw [red, ultra thick,->] (3.2,1.5)  -- (3.2,2.1) ;
\node [right, above right] at (0,0) { \tiny CDF Run II, 2.1 \invfb };
\node [right, above right] at (3.8,3.8) { \includegraphics[width=1cm]{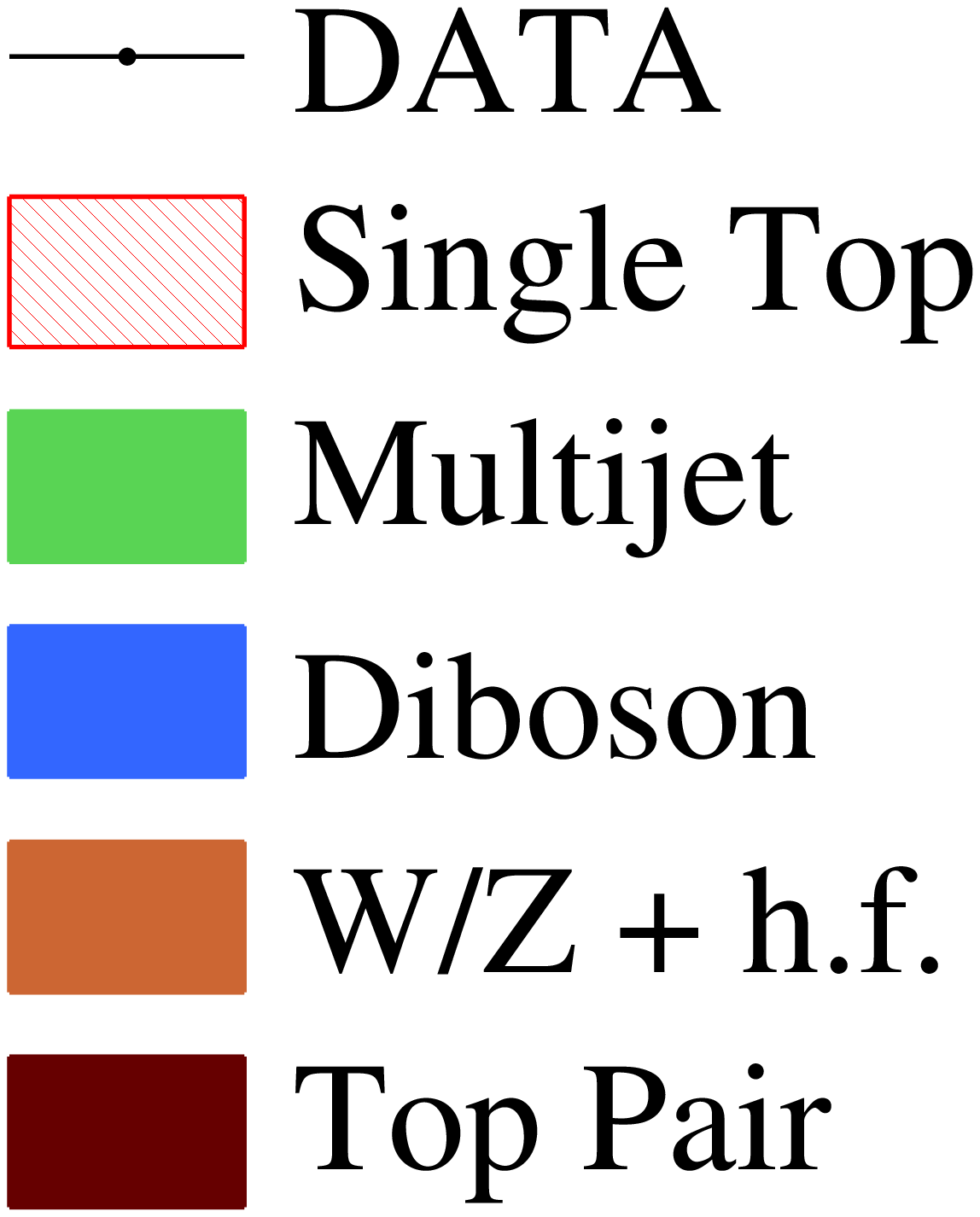}};

\node [inner sep=0pt,above right] at (0,-4) {\includegraphics[width=4.9cm]{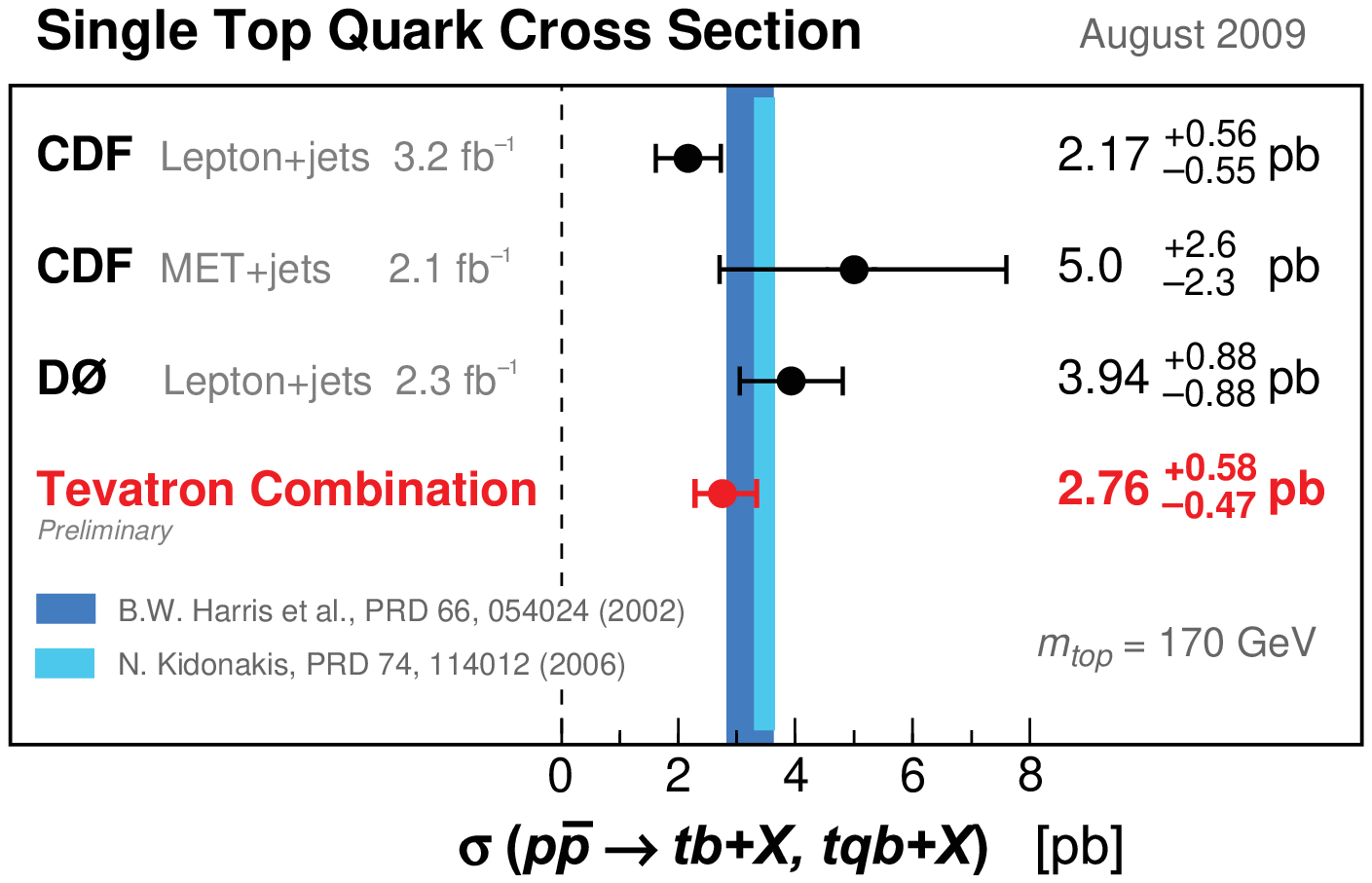}};
\draw [red,thick] (.2,-1.95) rectangle (4.7,-1.65);
\end{tikzpicture} 
}
\subfloat[SM Higgs]{\label{fig:higgsMETbb}
\begin{tikzpicture}
\node [inner sep=0pt,above right] at (0,0) {\includegraphics[width=5cm]{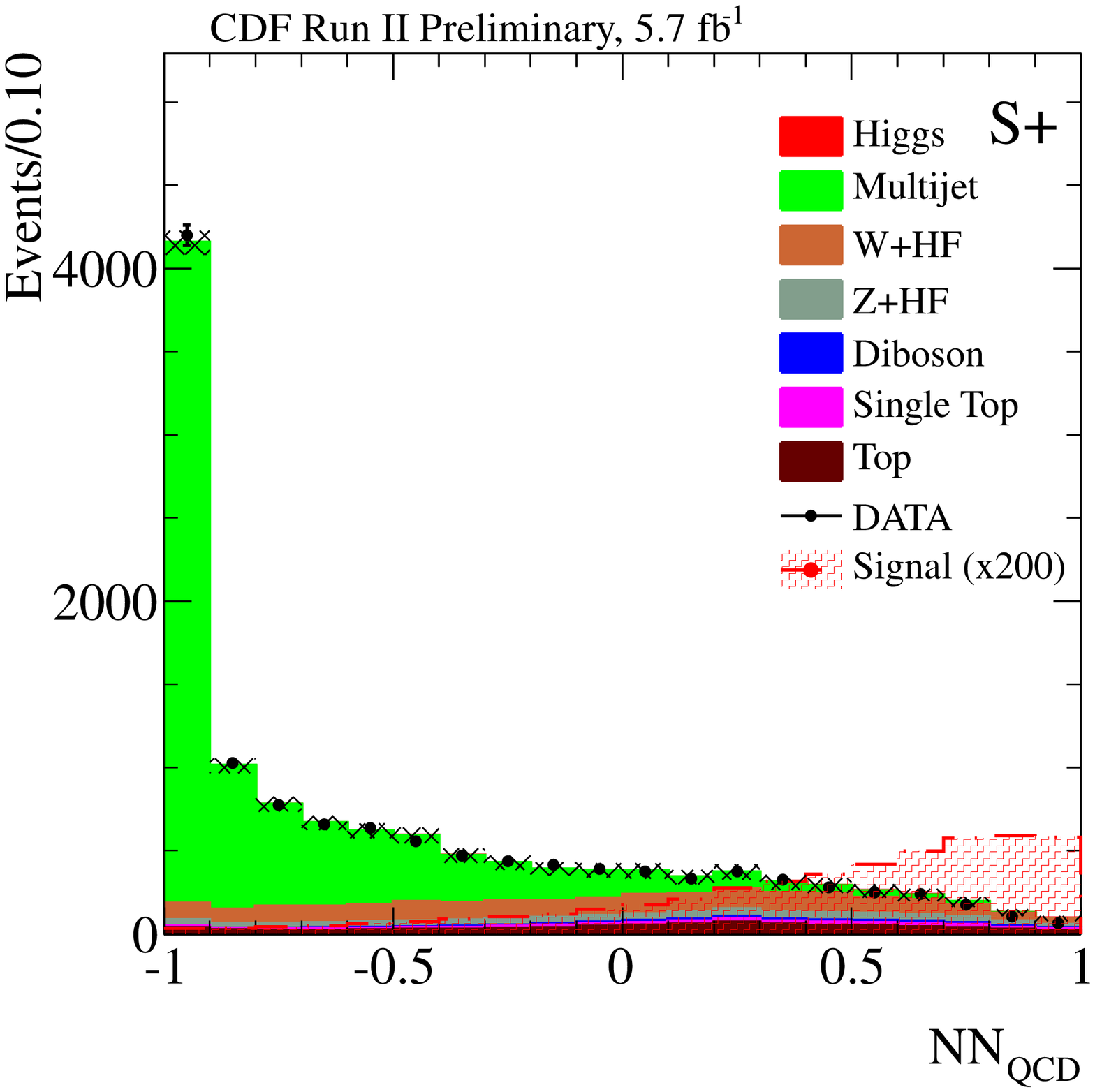}};
\node [inner sep=0pt,above right] at (1.7,1.6) {\includegraphics[width=3cm]{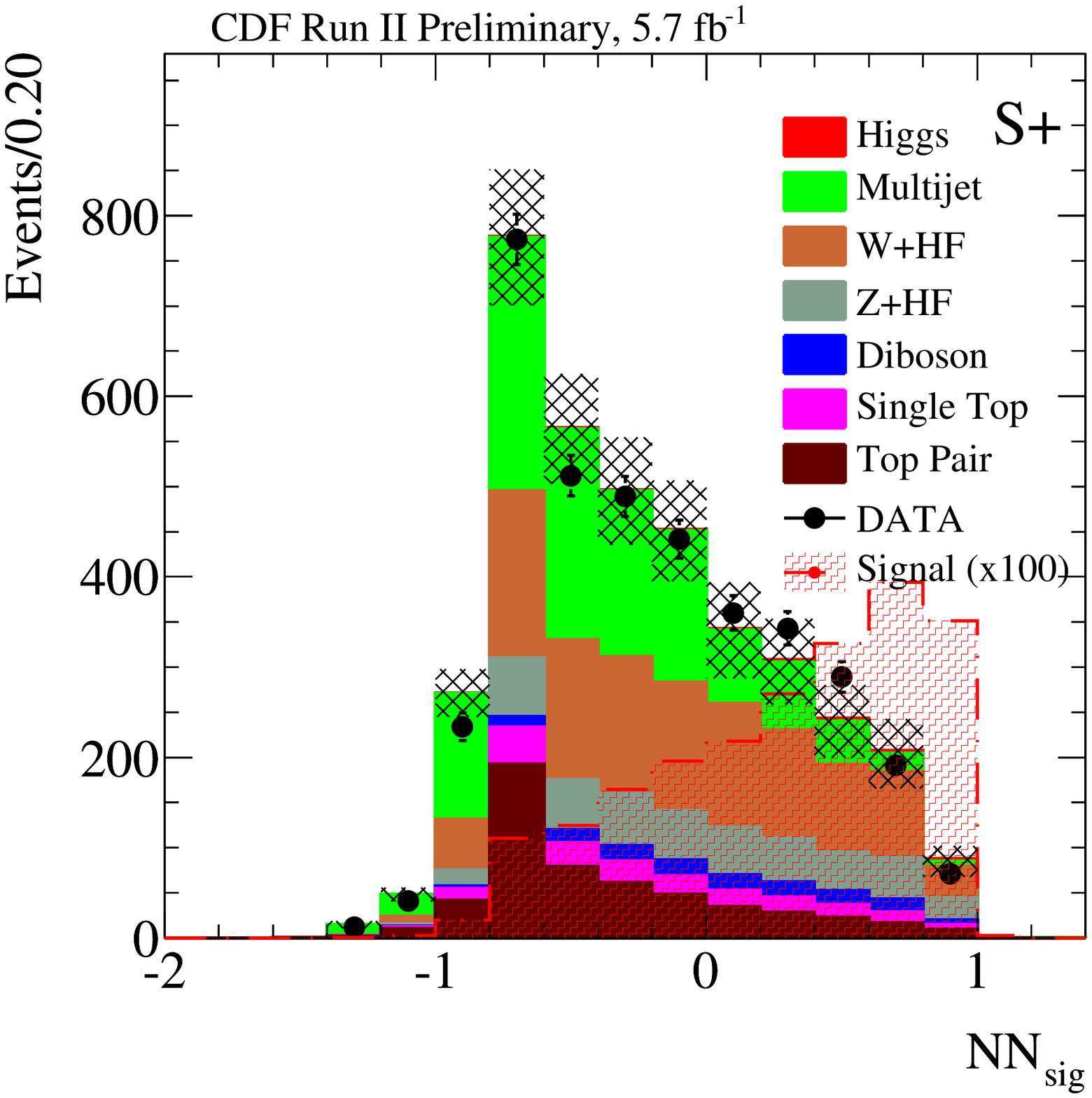}};
\draw [red, thick] (2.2,0.7) -- (2.2,1.9);
\draw [red,thick] (3.5,1.2) ellipse (1.3 and .5);
\draw [red, ultra thick,->] (3.4,1.4)  -- (3.4,2.0) ;

\node [inner sep=0pt,above right] at (0,-4) {\includegraphics[width=5.5cm]{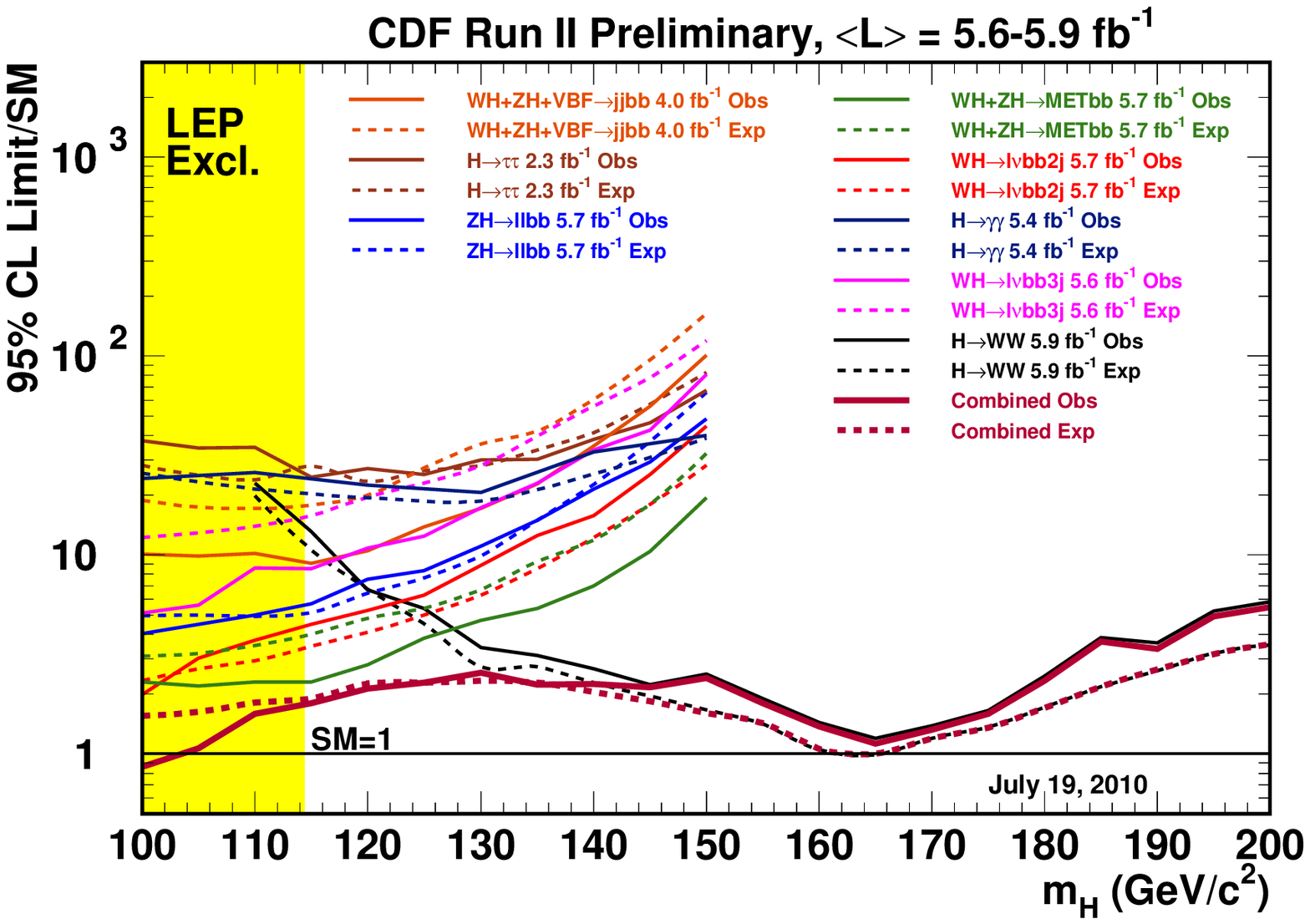}};
\draw [red,thick] (3.3,-0.75) rectangle (4.9,-0.5);
\draw [red,thick,rotate=25,yshift=-3.75cm] (1,.7) ellipse (1 and .25);
\end{tikzpicture}
}

\end{center}
\caption{Three analyses using a neural network to efficiently reject the QCD multi-jet background and another one to discriminate the signal from the remaining background: (a) Top pair cross-section, (b) EW single top cross-section, and (c) limit on SM Higgs. }
\end{figure}

The \met+$b$-jets signature is one of the most sensitive for probing a low mass SM Higgs boson, and an important component of the CDF and Tevatron combination~\cite{Potamianos:2010vp}. This search is challenging due to the tiny signal. We tune the analysis to $ZH\to\nu\nu b\bar{b}$ and $WH\to\mlep\nu b\bar{b}$~\cite{Aaltonen:2009jg,cdfNote:10212}. Analyzing 5.7\invfb\ of CDF data, we set a limit below 5 times the standard model prediction for a Higgs mass up to 135\gevcc, and at 3 times for $m_H=115$\gevcc\ (Figure~\ref{fig:singleTopMETbb}).

\vspace*{-2mm}
\section{Summary and future prospects}
\vspace*{-2mm}

The background rejection technique presented here yields similar performance to lepton identification and allows to exploit the large acceptance of the \met+$b$-jets signature. We have presented several analyses using this innovative technique to isolate small signals. Further steps include relaxing the pre-selection requirements (lower \met\ and jet energy cuts) to gain in acceptance while still rejecting the large backgrounds. We then plan to measure the diboson production cross-section in this signature.
\vspace*{-2mm}
\section*{Acknowledgments}
\vspace*{-2mm}
We would like to thank the organizers of the XLVI Rencontres de Moriond for an excellent conference and the CDF collaboration for making these results possible.

\end{document}